\documentclass[superscriptaddress,
reprint,
 amsmath,amssymb,
 aps,
]{revtex4-2}

\usepackage{graphicx}
\usepackage{graphics}
\usepackage{subfigure}
\usepackage{epstopdf}
\usepackage{dcolumn}
\usepackage{bm}
\usepackage{amsmath}
\usepackage{breqn}
\DeclareMathOperator{\Tr}{Tr}
\usepackage{amssymb}
\usepackage{mathrsfs}
\usepackage{amsfonts}
\usepackage{physics}
\usepackage{mathtools}
\usepackage[export]{adjustbox}

\makeatletter
\def\cat@comma@active{\catcode`\,12}%
\makeatother

\begin{document}
\preprint{APS/123-QED}
\title{Long range entanglement in quantum dot systems under the Fermi-Hubbard approach}

\author{Sanaa Abaach}
\email{sanaa.abaach@um5r.ac.ma}
\affiliation{ESMaR, Faculty of Sciences, Mohammed V University in Rabat, Morocco.}

\author{Mustapha Faqir}
\email{mustapha.faqir@uir.ac.ma}
\affiliation{Université internationale de Rabat, Aerospace engineering school, LERMA lab, Morocco.}

\author{Morad EL Baz}
\email{morad.elbaz@um5.ac.ma}
\affiliation{ESMaR, Faculty of Sciences, Mohammed V University in Rabat, Morocco.}

\date{\today}

\begin{abstract}
In the present paper we are interested in analyzing the pairwise entanglement in quantum dots, as ququart systems, naturally described by the Fermi-Hubbard model. Using the lower bound of concurrence we show the effect of the Coulomb interaction on the pairwise entanglement and creating entanglement within the pairs. Specifically, it is shown that the range of entanglement can be extended to the third neighboring site for a system size of $L=4$, while for $L>4$ the range could be created and extended to the third neighboring site by means of the Coulomb interaction. A rigorous description of the pairs is given in terms of a local half filled state associated to each pair with an electron number $N=2$ and a spin $S=0$. A thorough study of this state provides a proper explanation related to the pairwise entanglement, namely its amount and its behavior under the effect of the Coulomb interaction as well as the system's size.
\end{abstract}

\maketitle


\section{\label{sec: section1}Introduction}

As a swiftly expanding and cross-disciplinary topic, quantum entanglement has been a subject of various studies that attracted many leading theorists and experimentalists from physics, computer science, as well as electronic engineering in recent years. Mostly due to its non-local \cite{ref1} connotation, this primarily intriguing feature of quantum mechanics is regarded as a precious resource and a key ingredient in many striking achievements that have been witnessed in the last decade in quantum communication and information processing \cite{ref2,ref3, ref4}.

Over the past several years, there has been a heightened interest in quantum many body systems in favor of quantum information and \textit{vice-versa} \cite{ref5}. This interest has been triggered as a consequence of the imperative need to enhance the understanding of the physics of many body systems for the purpose of operating the basic unit of quantum information, \textit{i.e.}, the qubit, as well as for building scalable devices designed to implement quantum information tasks in a very accurate and controlled way. Among the most scalable and time-coherent platforms, dedicated to implement quantum information schemes, are quantum dot (QD) systems \cite{ref6,ref7,ref8}. QDs are semiconductor nanocrystals, often referred to as artificial atoms, where the carriers' motion is quantized in all the spatial directions giving rise to discrete energy levels (quantum confinement effect). Semiconductor QDs are naturally described by the Fermi–Hubbard (FH) model at the low temperature and strong Coulomb interaction regime \cite{ref9}. This property provides a simplified framework for the understanding of the behavior of QDs. As a matter of fact, the FH model is a simple approach that describes interacting spin- $ \frac{1}{2}$ fermions in many-body systems. The Coulomb repulsion interaction governing the fermions in this approach is one of the physical effects that dominates the electronic properties of QD systems \cite{ref10}. Hence, a better understanding of the Hubbard physics allows an accurate experimental tunability of the electron numbers in each dot, as well as other parameters, using gate voltages \cite{ref9}.

The ground state, of the FH model, is considered as a natural source of entanglement. Moreover, a closer look at the nature of entanglement in the FH model’s ground state, may reveal itself powerful to enhance the information. Indeed, the quantum states of such a model are \textit{ququart} states ($d=4$), giving rise to an enlarged Hilbert space. This could provide a larger information capacity and an increased noise resilience \cite{ref11} which could strongly promote the use of this model to carry out quantum information tasks. Nonetheless, increasing the dimension paves the way to the well known issue of entanglement quantification for mixed states in higher dimensional systems. In view of this fact, many questions are still open in the FH model and little is known about the details of entanglement, such as the behavior of pairwise entanglement as well as the range of entanglement in the weak interaction regime. In this regard, Li \textit{et al} \cite{ref13} have recently suggested a particular lower bound of concurrence (LBC) providing a good estimation of mixed states' multipartite entanglement in higher dimensions. Using this measurement we will show, for instance, that at weak interaction regime, the pairwise entanglement increases  with the growth of the coulomb interaction in contrast to the local entanglement \cite{ref23}.

In general, the most natural way of creating a large amount of entanglement between two or more parties, in low-dimensional systems, requires the presence of strong correlations. In most qubit systems with short-range interactions and with periodic boundary conditions, the entanglement between a pair of particles declines rapidly with the distance and could vanish even for distances larger than two sites such as the case of the Ising model with transverse field \cite{ref14}. It can also be restricted only to nearest neighbors in the Heisenberg model \cite{ref15}. From the point of view of quantum information, an even more appealing goal would be the ability to create a large amount of entanglement between distant and generally not directly interacting constituents. This can be conveniently exploited for an efficient implementation of spin chains based quantum information schemes \cite{ref16,ref17}. Moreover, it has been shown that the ground state of some spin chains with open boundary conditions and finite correlation length can supply large values of end-to-end entanglement between the end sites of the chain \cite{ref18}. In the present paper a characteristic case of an open array with size $L=4$ will provide an important end-to-end entanglement where the range of entanglement is well extended to the third neighbor, while for $L>4$ the range can be created and extended to the second and the third neighbors under the effect of the Coulomb interaction.

In this study, we will focus on finite-size systems as appropriate systems for numerical treatment that can be extrapolated to larger size lattices. Furthermore they are easily controlled and manipulated experimentally in nanotechnological applications. A pivotal analysis of our results will be presented in terms of a local half filled state (LHFS) describing each pair with two electrons. A complete knowledge about the LHFS and its associated probability will provide primary details regarding the amount of entanglement and its behavior under the effect of the Coulomb interaction as well as the system's size. The principal corner stones of our work are presented in Sec. \ref{sec: section2}: on one hand the one-dimensional FH model which provides an adequate approach for the description of QD systems and the LBC as a measure of pairwise entanglement on the other. The aforementioned measure allows us in Sec. \ref{sec: section3} to reveal the range of entanglement as well as the behavior of pairwise entanglement which proves to be distinct from that of the well-known local entanglement. Focusing on the smallest size that we simulated, $L=4$, a detailed explanation of the pairwise entanglement properties will be presented in Sec. \ref{sec: section4}, showing the presence of two competing effects responsible for the increase in pairwise entanglement: the mixing effect on one hand and the inherent effect associated to the LHFS on the other hand. Finally, our conclusions and perspectives are presented in Sec.\ref{sec: section5}.

\section{\label{sec: section2} Model and Formalism}

\subsection{Fermi-Hubbard Model}
\quad An array of QDs, can be modeled by the one-dimensional FH approach \cite{ref9,ref19,ref20}. Assuming that the hopping is bounded by the nearest-neighbor lattice sites, the simplest expression of the Hamiltonian corresponding to the model \cite{ref21} is formulated as follows
\begin{equation}
H = -t \sum_{i,\sigma} \left( c_{i,\sigma}^{\dag} c_{i+1,\sigma}+ c_{i+1,\sigma}^{\dag} c_{i,\sigma} \right) + u \sum_{i} n_ {i,\uparrow} n_{i,\downarrow},    
\label{ham}
\end{equation}
where $c_{i,\sigma}^{\dag}$ and $c_{i,\sigma}$ are, respectively, the creation and annihilation operators that describe the tunneling of electrons between the neighboring sites. $t$ is the hopping amplitude and $\sigma=\{ \uparrow,\downarrow \}$ indicates spin-up or spin down electron, whereas $u$ is the on-site electron-electron Coulomb interaction. We assume that only the $s$-orbital is allowed to electrons in each QD; accordingly each QD is able to hold up to two electrons with opposite spins in compliance with the Pauli exclusion principle. Thereby electrons have four possibilities in occupying a single site: $\ket{0}$, $\ket{\uparrow}$, $\ket{\downarrow}$ and $\ket{\uparrow\downarrow}$ (standing respectively for: no electron, one electron having a spin up, one electron having a spin down and two electrons).
Indeed, when the repulsion interaction $u$ within the sites is too strong, the tunneling of electrons $t$ is blocked leading to a clear observation of the quantum confinement effect in the FH model. This physical picture is analogous to the formation of potential barriers between the sites that prohibits electrons from tunneling outside. Experimentally, the creation of such barriers is made by modulating potentials, using gate electrodes, in order to control the tunneling of electrons between the quantum dots \cite{ref9}. In the following we will consider the dimensionless quantity $U=u/t$ as the main parameter in the model.

\subsection{Lower bound of concurrence (LBC)}

The absence of a proper measure of entanglement for systems beyond the $2 \times 2$ and $2 \times 3$ dimensions, forces us to settle for a measure of the lower bound of entanglement. For an arbitrary bipartite system of dimension $d\times d$, the concurrence $C(\rho_{ij})$ \cite{ref13} satisfies
\begin{equation}
  \tau_{2}( \rho_{ij})=\frac{d}{2(d-1)}\sum_{\alpha}^{\frac{d(d-1)}{2}}\sum_{\beta}^{\frac{d(d-1)}{2}} C_{\alpha\beta}^{2}\le C^{2}(\rho_{ij}),
  \label{pairlow}
\end{equation}
where
\begin{equation}
C_{\alpha\beta}=max\{0,\lambda_{\alpha\beta}^{(1)}-\lambda_{\alpha\beta}^{(2)}-\lambda_{\alpha\beta}^{(3)}-\lambda_{\alpha\beta}^{(4)} \}. \\
\end{equation}
In our case, $ \rho_{ij}$ is the pairwise density matrix of a four-level system (i.e., ququarts) for concreteness we take  $i<j$. $\lambda_{\alpha\beta}^{(m)} $ are the square roots of the non-zero eigenvalues of the non-Hermitian matrix $ \rho_{ij}\tilde{\rho}_{(ij)\alpha\beta} $ such that 
$ \lambda_{\alpha\beta}^{(m)}> \lambda_{\alpha\beta}^{(m+1)} $ for $ m=1, 2, 3 \; \hbox{or} \; 4$ and
\begin{equation}
   \tilde{\rho}_{(ij)\alpha\beta}=(G_{\alpha}\otimes G_{\beta})\rho_{ij}^{*}(G_{\alpha}\otimes G_{\beta}).
\end{equation}
$G_{\alpha}$ is the $\alpha^{\text{\tiny th}}$ element of the group $SO(d)$ spanned by $\frac{d(d-1)}{2}$ generators. $G_{\beta}$ is defined similarly since the two subsystems have the same dimension $d$.

The left hand side of the inequality in (\ref{pairlow}) can be effectively computed, providing thus a solid lower bound of concurrence (LBC) and entanglement.

Although being deficient in granting the entire details about the amount of entanglement, this LBC imparts precious information that is not obtainable otherwise. The efficiency of the LBC is manifested in the fact that it can detect mixed entangled states with a positive partial transpose \cite{ref22} and that for fully separable multipartite state it is equal to zero.

\section{\label{sec: section3} The range of entanglement and the size effect}

\begin{figure*}
  \centering
  \subfigure[]{\includegraphics[scale=0.28]{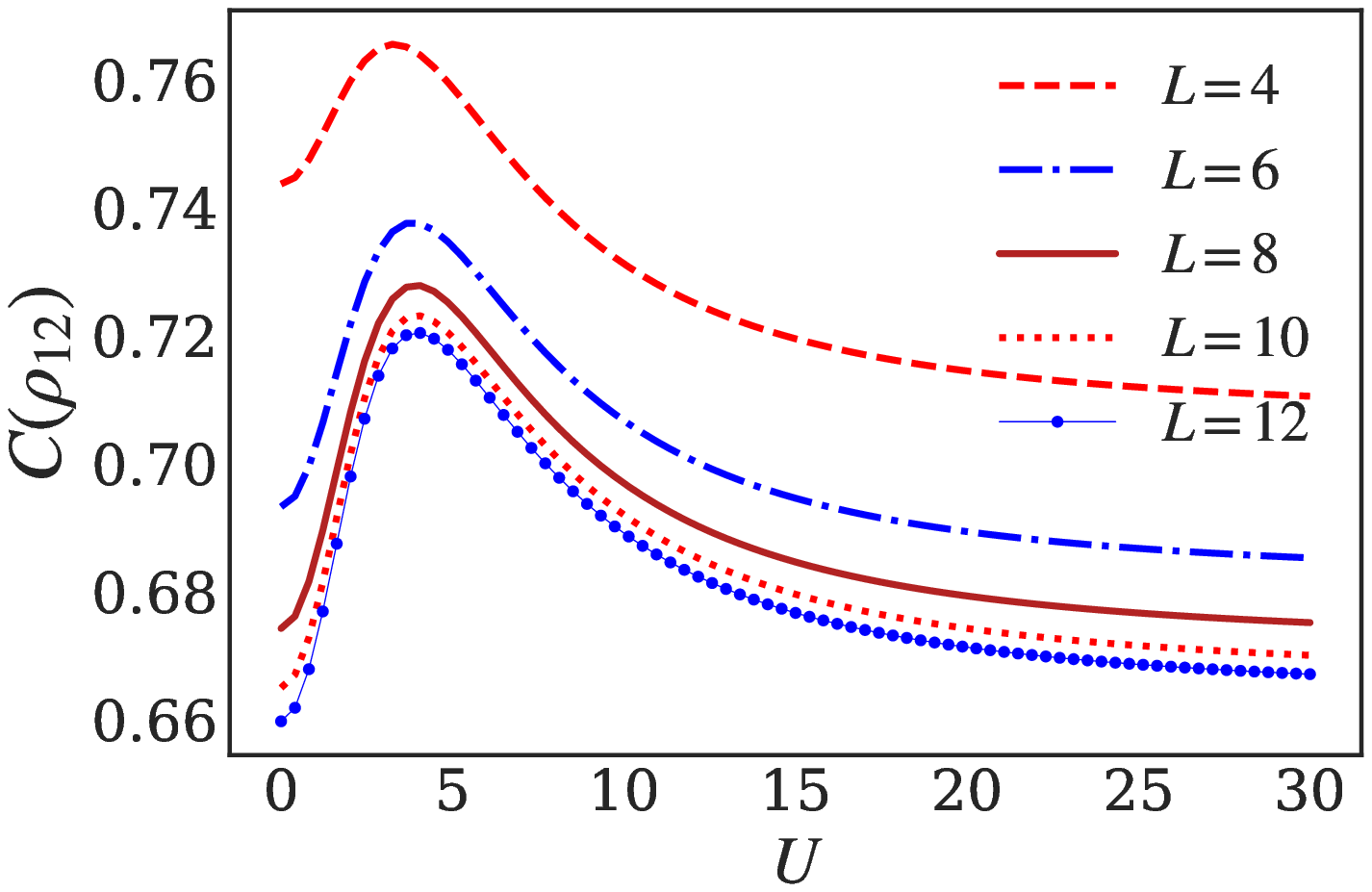}
  \label{fig:img1a}}\quad
  \subfigure[]{\includegraphics[scale=0.285]{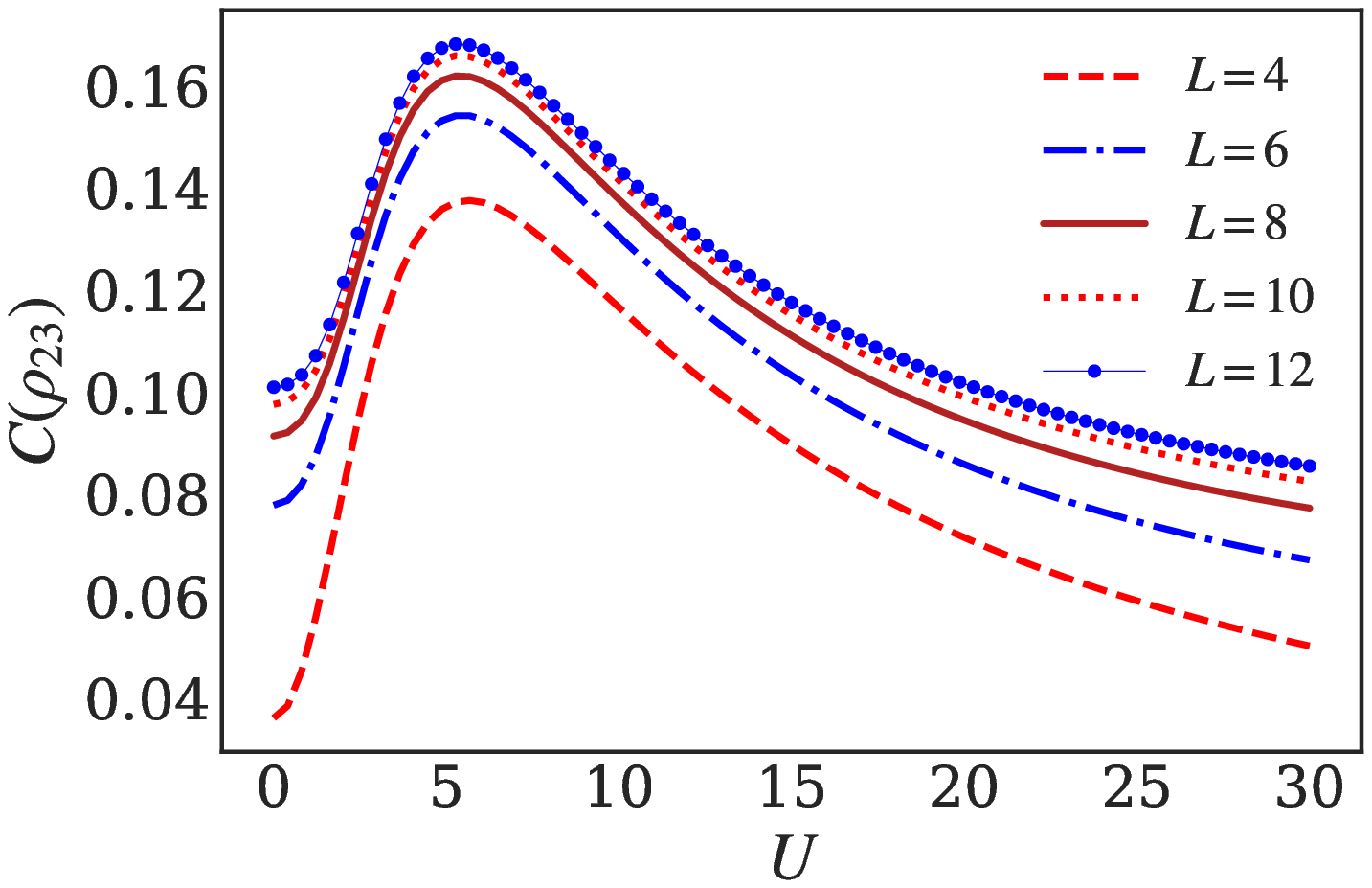}
  \label{fig:img1b}}\quad
  \subfigure[]{\includegraphics[scale=0.28]{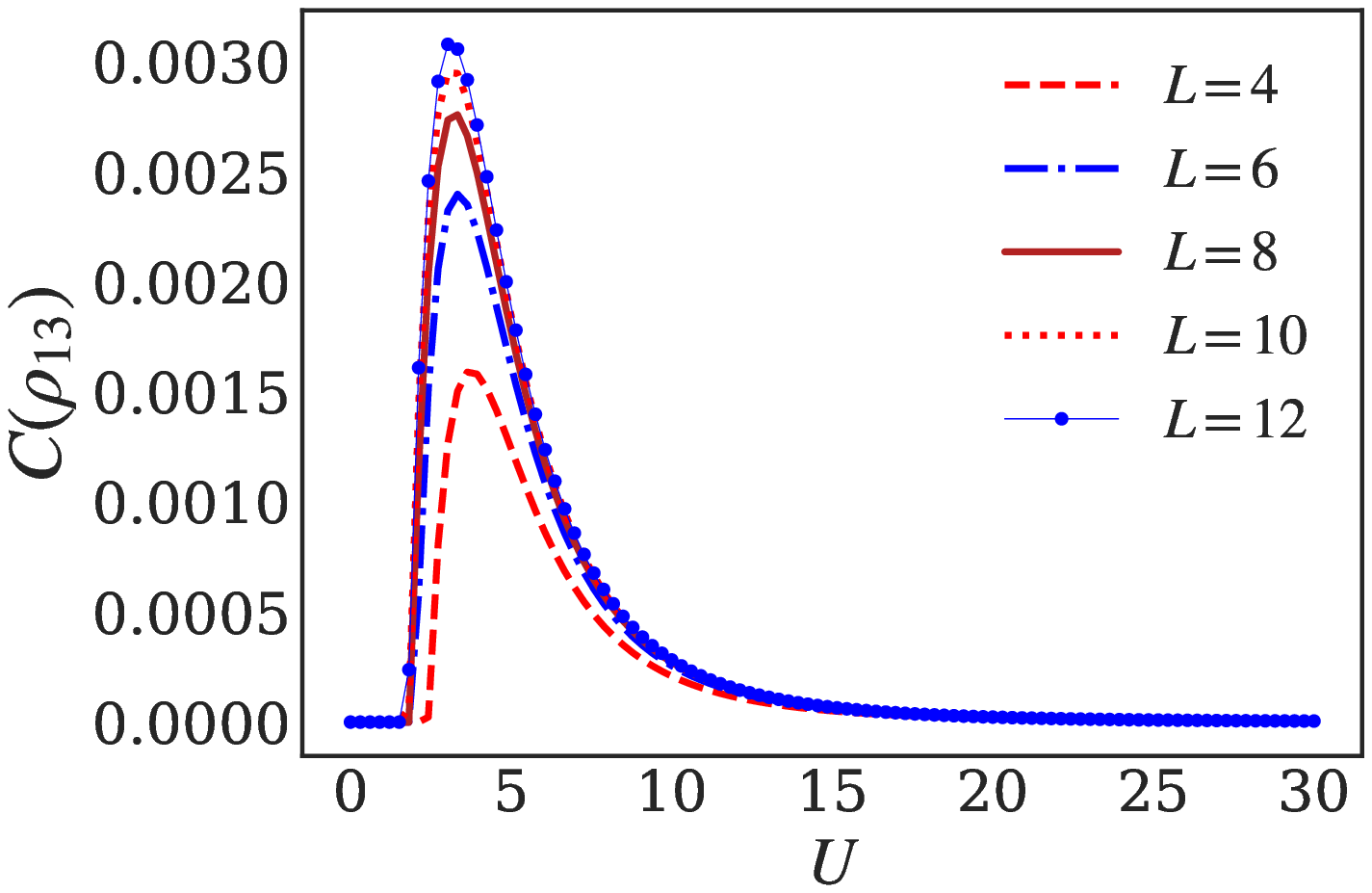}
  \label{fig:img1c}}
  \subfigure[]{\includegraphics[scale=0.28]{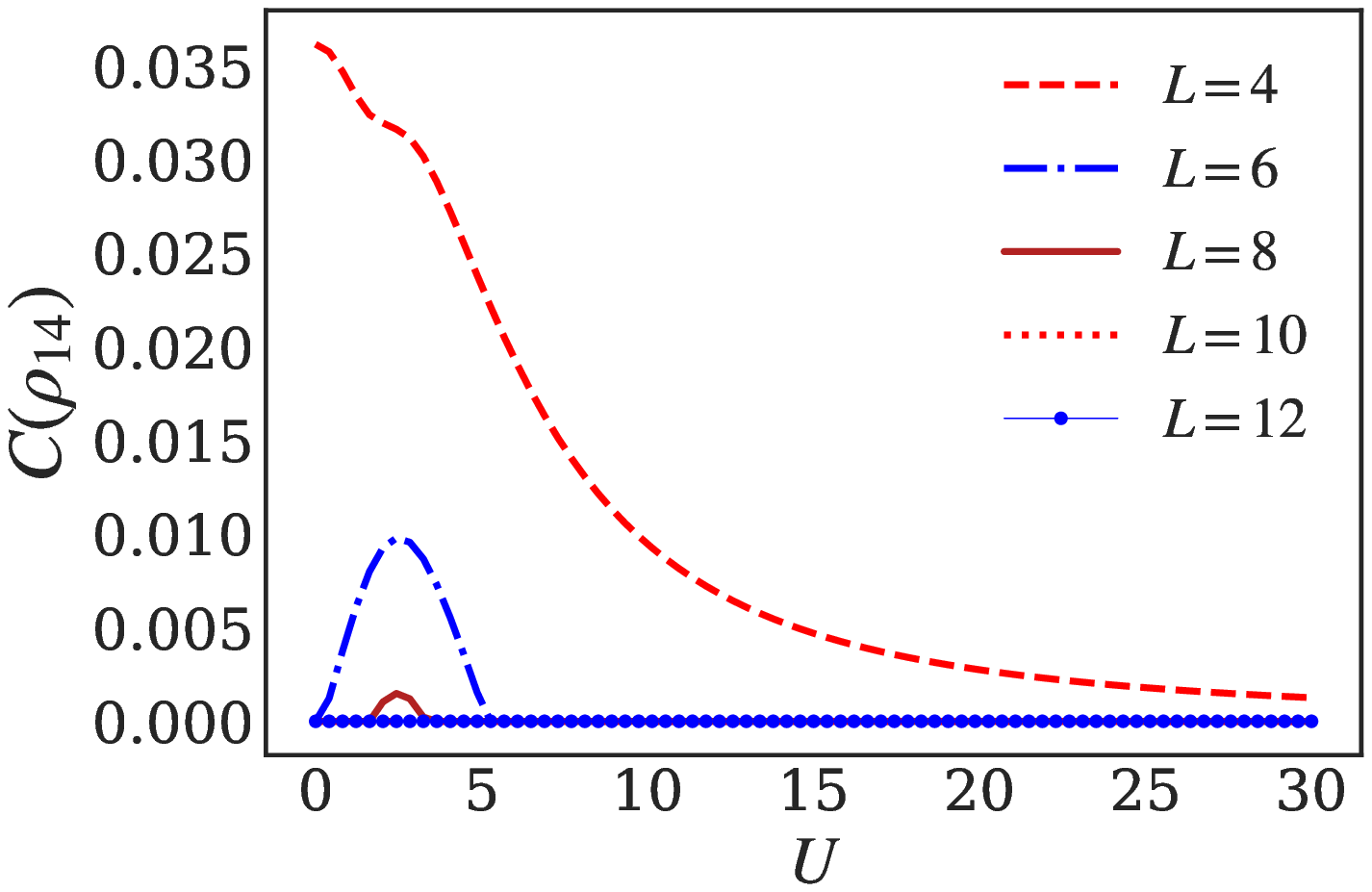}
  \label{fig:img1d}}
  
  \caption{\textbf{Lower bound of concurrence.} The LBC, Eq.\eqref{pairlow}, associated to the pair (a) $\rho_{12}$, (b) $\rho_{23}$, (c) $\rho_{13}$ and (d) $\rho_{14}$ as a function of $U$ for various system sizes up to $L=12$.}
  \label{fig:img1}
\end{figure*}

In this section, we analyze the pairwise entanglement behavior, under the effect of the Coulomb interaction $U$ in a finite size array of QDs, formally described by the FH Model. Furthermore the range of entanglement under the system's size effect is investigated. This will establish the peculiarities of this type of entanglement compared to the single site entanglement.
FIG. \ref{fig:img1}, gives us some insights in this regard. An initial increase of $U$, in the weak coupling regime, could give rise to an appreciable growth in the pairwise entanglement in contrast to the local entanglement that decays instantly with $U$ \cite{ref23}.  Since entanglement is a crucial ingredient for quantum information tasks, any improvement or increase in this resource will make a prominent contribution in quantum information. In this regard, $U$ proves its potency in rising and, more importantly, creating entanglement within the pairs and this will play a significant role in the direction that employs spin chains as quantum channels \cite{ref24}. 

Since intuitively the presence of direct and strong correlations provide great amounts of entanglement, the adjacent sites, FIG.\ref{fig:img1a} and FIG.\ref{fig:img1b}, for instance, should reveal a larger amount of entanglement compared to the distant sites FIG.\ref{fig:img1c} and FIG.\ref{fig:img1d} at $U=0$. Indeed the previous behavior is in agreement with the monogamy of entanglement which broadly stipulates that if two systems are strongly entangled with each other, then each of them cannot be entangled very much with other systems. In addition to that, when increasing $U$ while staying in the weak coupling regime, the entanglement (LBC) grows considerably for the adjacent sites but insubstantially for the distant sites (except for $\rho_{14}$ when $L=4$). After reaching a maximal value, the LBC diminishes asymptotically to a non-zero value for nearest neighboring sites and to zero for distant sites. 
Furthermore from the point of view of the system's size, FIG.\ref{fig:img1} shows that as the system grows in size the pairwise entanglement $C(\rho_{ij})$ decays for all the pairs $\rho_{ij}$ ($i<j$) with $j$ being even but increases instead for all the pairs where $j$ is odd. Generally this can be interpreted by the fact that the pairs at the borders $\rho_{12}$ and $\rho_{L-1,L}$, are less correlated with the other sites according to the monogamy of entanglement as the pairs have to conserve the higher amount of entanglement, as shown in FIG.\ref{fig:img1a}. With the increase of the size the influence of the other sites becomes increasingly important, consequently the pairwise entanglement $C(\rho_{12})$ and $C(\rho_{L-1,L})$ at the borders decays with the system size. Since the pairs at the borders have an even $j$, their nearest pairs, \textit{i.e.}, $\rho_{23}$ and $\rho_{L-2,L-1}$ which necessarily have an odd index $j$, have to be more entangled with the border sites (\textit{i.e.} $1$ and $L$ respectively), according to monogamy of entanglement, as the system size increases. This, for instance, explains the increase of the pairwise entanglement $C(\rho_{13})$ and $C(\rho_{23})$ in FIG.\ref{fig:img1} the size $L$ of the system.

One of the most important aspects characterising spin chain systems, is their ability to distribute entanglement between distant parties. It is clearly noticed from FIG.\ref{fig:img1d} that at $U=0$, the range of entanglement is well extended to the third neighboring site for the smallest size system $L=4$, while for $4<L<10$ the range of entanglement is restricted to the nearest neighboring sites.  Nevertheless it is created and extended again to the third neighboring site with an optimal choice of $U$ in a specific interval, whereas with the size growth $L\geq 10$ the range of entanglement is restricted to the second neighbors again in a specific interval of $U$ (FIG.\ref{fig:img1}). Beyond that the pairwise entanglement vanishes for distances larger than single neighboring site when $U$ takes large values. 
Finding an appropriate interpretation for the increase in pairwise entanglement as a function of $U$, in the week coupling regime, can seem to be rather complicated as the question concerns the larger sizes. To circumvent this initially, in the next section, our study will be based on the smallest size systems $L=4$ where a detailed explanation of the aforementioned pairwise entanglement behavior can be easily analyzed and afterward generalized to larger system sizes.

\section{\label{sec: section4} Entanglement and the local half filed state}

\subsection{\label{sec: subsection3}The local entanglement}

For a chain with open boundary conditions, because of lacking translational invariance, the local entanglement (referring to the entanglement of a given site with the rest of the chain) is not expected to be the same for each site $i$. However, the mirror reflection symmetry translates into the following constraint $E(\rho_{i}) = E(\rho_{L-i+1})$, where $E(\rho)=-\Tr(\rho\log_{2}\rho)$ is the von Neumann entropy and
\begin{equation}
     \rho_{i}= v_{i} \ket{0} \bra{0} + s_{i\uparrow} \ket{\uparrow} \bra{\uparrow} + s_{i\downarrow} \ket{\downarrow}\bra{\downarrow} + d_{i} \ket{\uparrow \downarrow}\bra{\uparrow \downarrow}
\end{equation}
 with
\begin{align}
   d_{i}
   &=\Tr(n_{i\uparrow}n_{i\downarrow} \rho_{i}) =\langle n_{i\uparrow}n_{i\downarrow}\rangle,  &  s_{i\uparrow} &=\langle n_{i\uparrow}\rangle - d_{i},\nonumber \\
   s_{i\downarrow} &= \langle n_{i\downarrow}\rangle - d_{i},  &  v_{i}&= 1- d_{i} + s_{i\uparrow} + s_{i\downarrow}.
\end{align}
For the FH model at half filling, $ \langle n_{i\uparrow} \rangle=\langle n_{i\downarrow} \rangle=\frac{1}{2}$, $s_{i\uparrow}= s_{i\downarrow}=s_{i}=\frac{1}{2}-d_{i}$. Consequently, the corresponding von Neumann entropy is, 
\begin{equation}{\label{vonNeumann}}
    E(\rho_{i})=-2d_{i}\log_{2}d_{i} - 2(1/2 -d_{i})\log_{2}(1/2 -d_{i}).
\end{equation}

\begin{figure}
  \centering
 
  \subfigure[]{\includegraphics[scale=0.35]{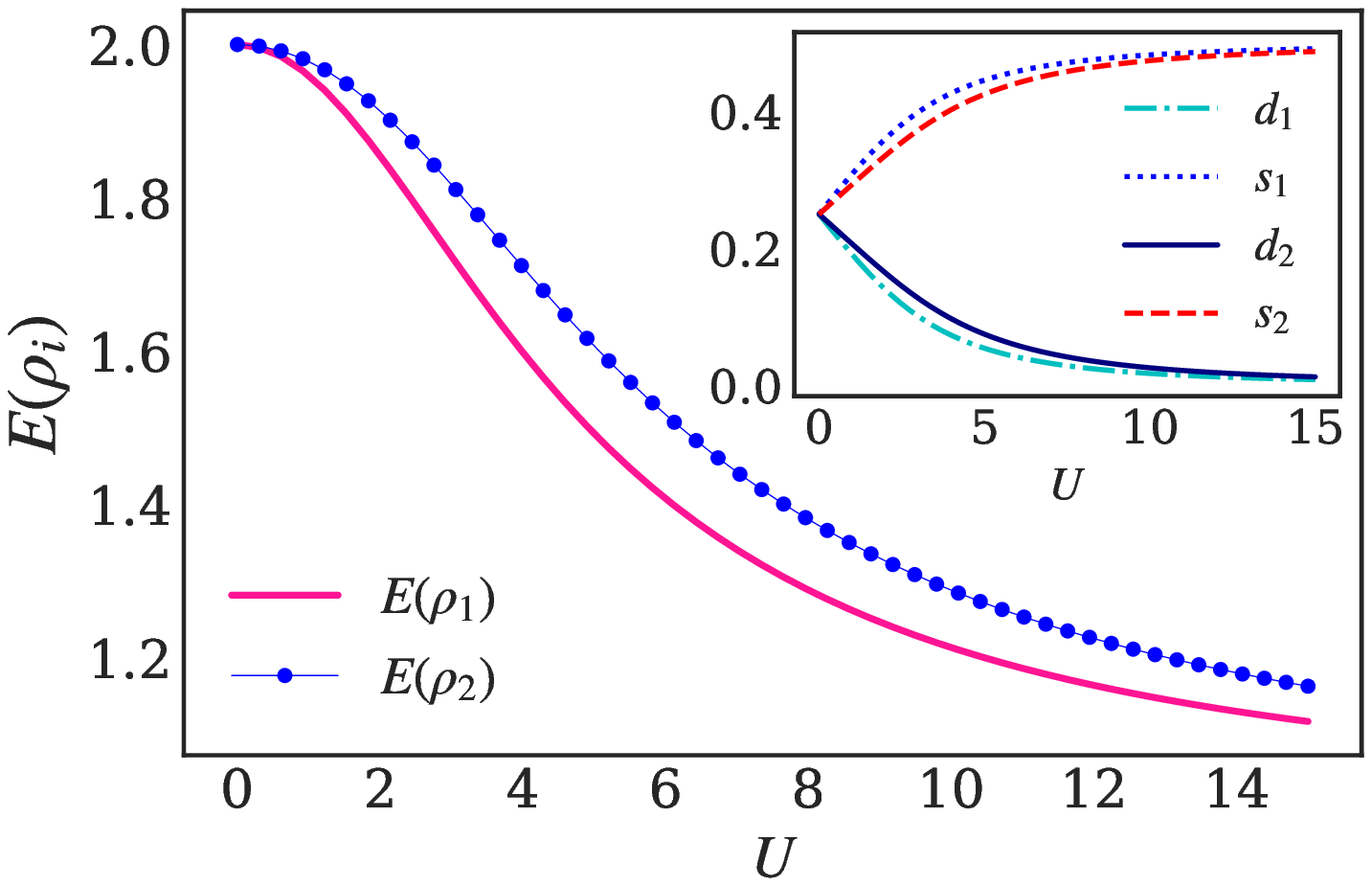}
  \label{fig:img3a}}\quad
   
  \subfigure[]{\includegraphics[scale=0.35]{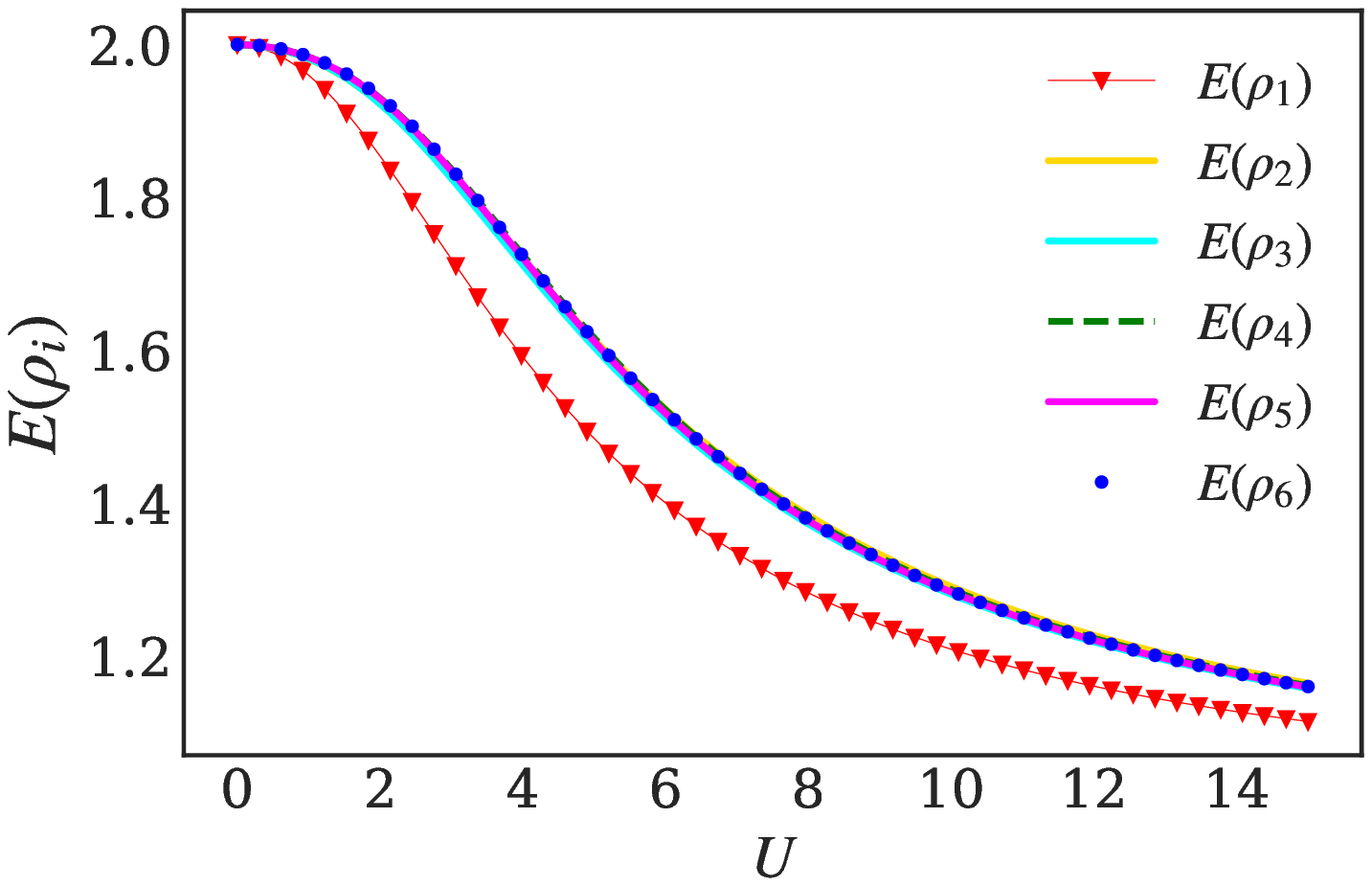}
  \label{fig:img3b}}\quad

  \caption{\textbf{Local entanglement.} (a) $E(\rho_{1}) = E(\rho_{L})$ and $E(\rho_{i})$ with $1<i<L $, Eq.\eqref{vonNeumann}, at the ends and the middle of the chain respectively as a function of $U$ for a system size (a) $L=4$ and (b) $L=12$. The \textit{upper right inset } in (a) shows the effect of $U$ on the single and double occupation probabilities $s_{i}$ and $d_{i}$, respectively at the end sites ($i=1,4$) and the middle sites ($i=2,3$).}
  \label{fig:img3}
\end{figure}

It can be seen from FIG.\ref{fig:img3} for $L=4$ (FIG.\ref{fig:img3a}) and $L=12$ (FIG.\ref{fig:img3b}), that while behaving similarly under the effect of $U$, the amount of local entanglement at the end sites, $E(\rho_{1}) = E(\rho_{L})$, is less than that in the inner sites, $E(\rho_{i})$ with $1<i<L $. It is quite clear that at $U=0$ the electrons can move freely in the array in such away that each site has the same probability for being singly occupied, $s_{i}$, doubly occupied, $d_{i}$, or empty, $v_{i}$. Focusing on the smallest size $L=4$, this is confirmed from the inset in FIG.\ref{fig:img3a} where at $U=0$, $s_{i}=d_{i} = \frac{1}{4}$, thus explaining why $E(\rho_{i})$ at the ends and the inner sites have the same value. Increasing $U$ the local entanglement at the ends exhibits a faster decrease compared to the entanglement in the middle of the chain $E({\rho_{1}}) = E({\rho_{4}})< E({\rho_{2}})=E({\rho_{3}})$. This can be easily explained following the same reasoning mentioned previously by noticing from FIG.\ref{fig:img3a} (the upper right inset), that as the repulsion interaction increases the sites at the ends favor the single occupancy ($s_{\uparrow}$ or $s_{\downarrow}$) more than the sites in the middle and therefore the correlations created at the end sites tend to degrade faster with the increase of $U$. The same behavior is consistently applied to large sizes, nevertheless, the sites in the middle retain the same amount of entanglement, as shown in FIG.\ref{fig:img3b}, considering the fact that the effect of the borders becomes negligible as we move away from the ends. Consequently the occupation probabilities $s_{i}$ and $d_{i}$ in this case becomes equal for all the middle sites, thus resulting in equal amounts of entanglement.

It is worth mentioning that, even though the on-site interaction $U$ is assumed to be the same in all the sites, the situation becomes equivalent to a chain with $U_{e}>U_{m}$ \cite{ref25} for finite $U$, where $U_{e}$ and $U_{m}$ denote the end and the middle on-site Coulomb interactions respectively. Generally The inequality $U_{e}>U_{m}$ expresses the fact that the occupation probabilities associated to the four configurations are not equal for all the sites. Notably, the occurrence of double occupancy at the end sites is more disfavored in comparison to that at the middle sites. This scenario is different for periodic chains where the local entanglement for all the sites is the same. 

At $U\to\infty$ The tunneling of electrons is blocked yielding thus to the confinement state, where each site confine one electron with spin up or spin down. At this stage the probability of the different occupation configurations turns equal at each site. As a result the entanglement at each site becomes the same in the confinement state. Understanding the behavior of the local entanglement behavior will partly explain the behavior of the pairwise entanglement discussed hereafter.

\subsection{Pairwise entanglement at $U=0$}

\begin{figure*}

  \centering
  \subfigure[]{\includegraphics[scale=0.43, valign=t]{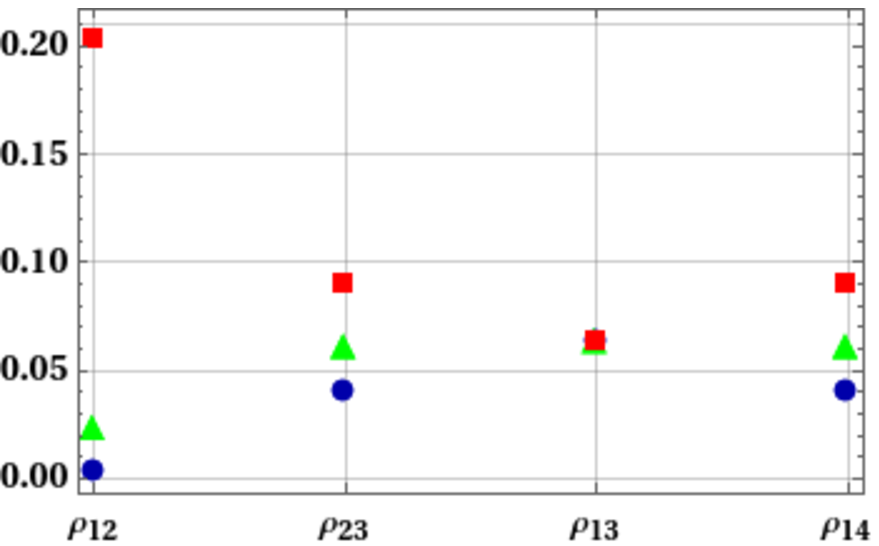}
  \label{fig:img4a}
   \vspace*{1cm}
  }
  \subfigure[]{\includegraphics[scale=0.54, valign=t]{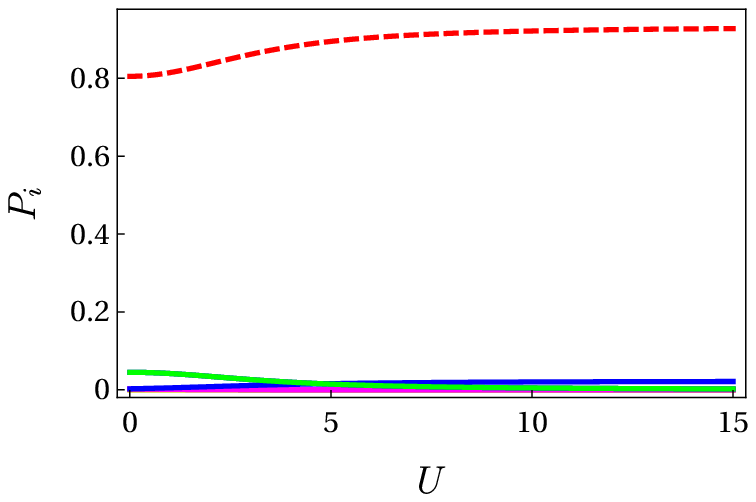}
  \label{fig:img4b}}\quad
  \subfigure[]{\includegraphics[scale=0.54, valign=t]{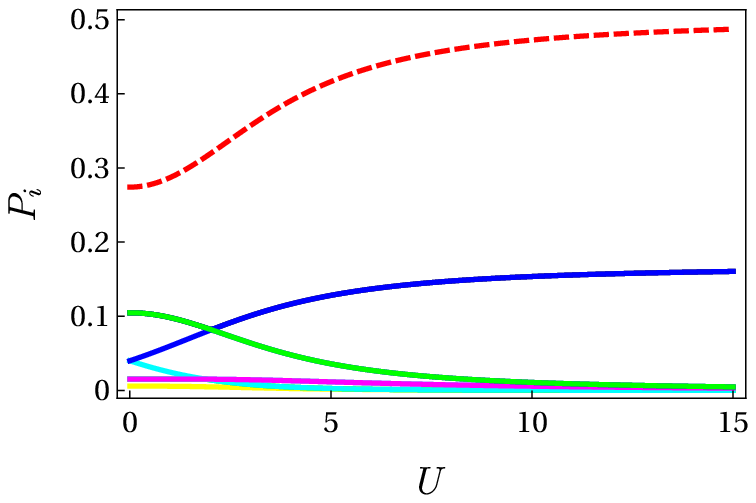}
  \label{fig:img4c}}\quad
  \subfigure[]{\includegraphics[scale=0.54, valign=t]{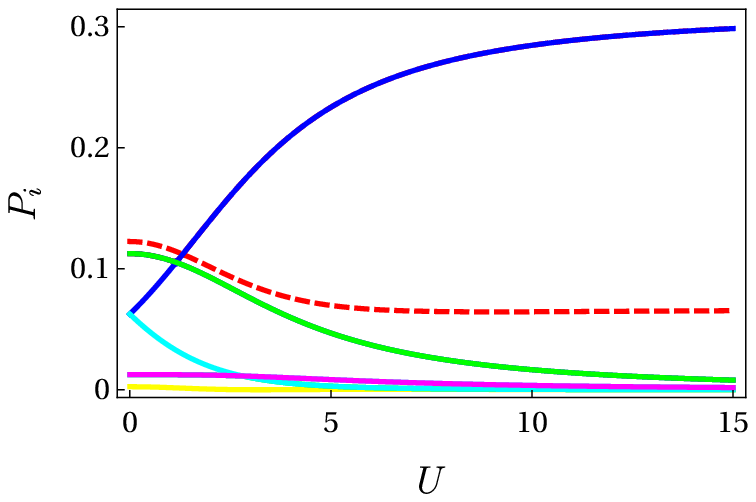}
  \label{fig:img4d}}

  \caption{(a) The probabilities corresponding to the sixteen basis states for each pair at $U=0$ with $L=4$. The basis states with $N=2$ and $S=0$ are marked by Red squares for, those with $N=\{1,3 \}$ and $S=\pm \frac{1}{2}$ by green triangles while the blue disks display basis states with $N=\{0,4\}$, $S=0$ and $N=2$, $S=\pm 1$. (b), (c), (d): The probabilities $P_{k}$ associated to the $4^2$ eigenstates $\ket{\psi_{k}}$ of $\rho_{12}$, $\rho_{23} = \rho_{14}$ and $\rho_{13}$ respectively. $P_{LHFS}$ associated to the LHFS is marked by a dashed line.}
  \label{fig:img4}
   
\end{figure*}

For the purpose of analyzing the pairwise entanglement behavior, we have to discuss generally the state characterising the pairs at $U=0$ and without loss of generality we consider $L=4$. In that case, the half filed single band array is defined by a fixed total number of electrons, $4$, and a null total spin. However the situation is different for the local pairs where a given (odd or even) number of electrons $N=\{0,1,2,3,4\}$ can be associated to each pair with a specific total spin $S=\{-1, -\frac{1}{2}, 0, \frac{1}{2}, 1\}$. Henceforth the state describing each pair is represented by a mixture 
\begin{equation}
\label{rhoij} 
    \rho_{ij}= \sum_{k} P_{k} \ket{\psi_{k}}\bra{\psi_{k}},
\end{equation} 
where $\ket{\psi_{k}}$ are the possible states corresponding to a given number $N$ and a total spin $S$ defined above.

As previously outlined, at $U=0$ each site has the same probability $\frac{1}{4}$ to be in one of the four possible states $\ket{\uparrow}$, $\ket{\downarrow}$, $\ket{\uparrow \downarrow}$ or the vacuum state $\ket{0}$. Nevertheless, the different pairs are described differently and they exhibit dissimilarity in the occupation probabilities associated with the sixteen ($4^2$) basis states of each pair. In FIG.\ref{fig:img4a}, it's remarked that the pairs of neighboring sites located at the ends of the array, $\rho_{12}=\rho_{34}$, have a relatively high probability to be in the states $\ket{\uparrow, \downarrow}$, $\ket{\downarrow , \uparrow}$, $\ket{\uparrow \downarrow, 0}$ and $\ket{0, \uparrow \downarrow}$, \textit{i.e.,} states with $N=2$ and $S=0$, compared to other states. The other sites exhibit a different behavior and as such, it becomes obvious that the pairwise entanglement cannot be the same between all pairs at $U=0$ contrary to the local entanglement. 

It is worth mentioning that all the states $\ket{\psi_{k}}$ in the decomposition (\ref{rhoij}) are degenerate due to particle number conservation and spin symmetry, except the state with $N=2$ and $S=0$ which is given by $(\alpha \ket{\uparrow, \downarrow} + \beta \ket{ \downarrow , \uparrow} +\gamma \ket{\uparrow \downarrow , 0} + \delta \ket{0, \uparrow \downarrow})$, hereafter referred to as the local half filed state, LHFS. Furthermore, for all the pairs at $U=0$, it is the state that possesses the highest probability in contrast to the others as reflected in FIG.\ref{fig:img4}. Given the fact that the LHFS is the dominant state as well as highly entangled state (maximally entangled for $\lvert  \alpha \rvert= \lvert \beta \rvert=\lvert  \gamma \rvert= \lvert  \delta \rvert = \frac{1}{2}$ at $U=0$) in the mixture (\ref{rhoij}), the behavior of the pairwise entanglement will be essentially dominated by the LHFS entanglement behavior itself. Henceforth, in the following the pairwise entanglement behavior will be explained basing on the LHFS.

In FIG.\ref{fig:img4b}, it is shown that the pair $\rho_{12}$ has a very high probability, that exceeds 0.8, to be in the LHFS. This is expected because as we have pointed out, the states $ \ket{\uparrow, \downarrow}$ , $\ket{ \downarrow , \uparrow}$ , $\ket{\uparrow \downarrow , 0}$ and $\ket{0, \uparrow \downarrow}$ are the most favored states for the pair $\rho_{12}$ (FIG.\ref{fig:img4a}), followed by the pair $\rho_{23}$, $\rho_{14}$ and finally $\rho_{13}$. As the mixture is generally dominated by the LHFS, it is evident that the higher the probability for being in this state, the higher the pairwise entanglement will be. This is what explains the pairwise entanglement at $U=0$ observed in FIG.\ref{fig:img1}, where, considering $L=4$ for illustration purposes, the highest amount of entanglement goes to the pair $\rho_{12}$ followed by $\rho_{23}$, $\rho_{14}$ and finally $\rho_{13}$. The same reasoning applies for $L>4$, where the amount of pairwise entanglement $C(\rho_{ij})$ depends primarily on the quantity $P_{LHFS}$ (the probability for a pair to be in the LHFS), in such a manner that, for a fixed $L$, the higher $P_{LHFS}$, the higher the pairwise entanglement $C(\rho_{ij})$ is and \textit{vice versa}. 

\begin{figure}
  \centering
  \subfigure[]{\includegraphics[scale=0.35]{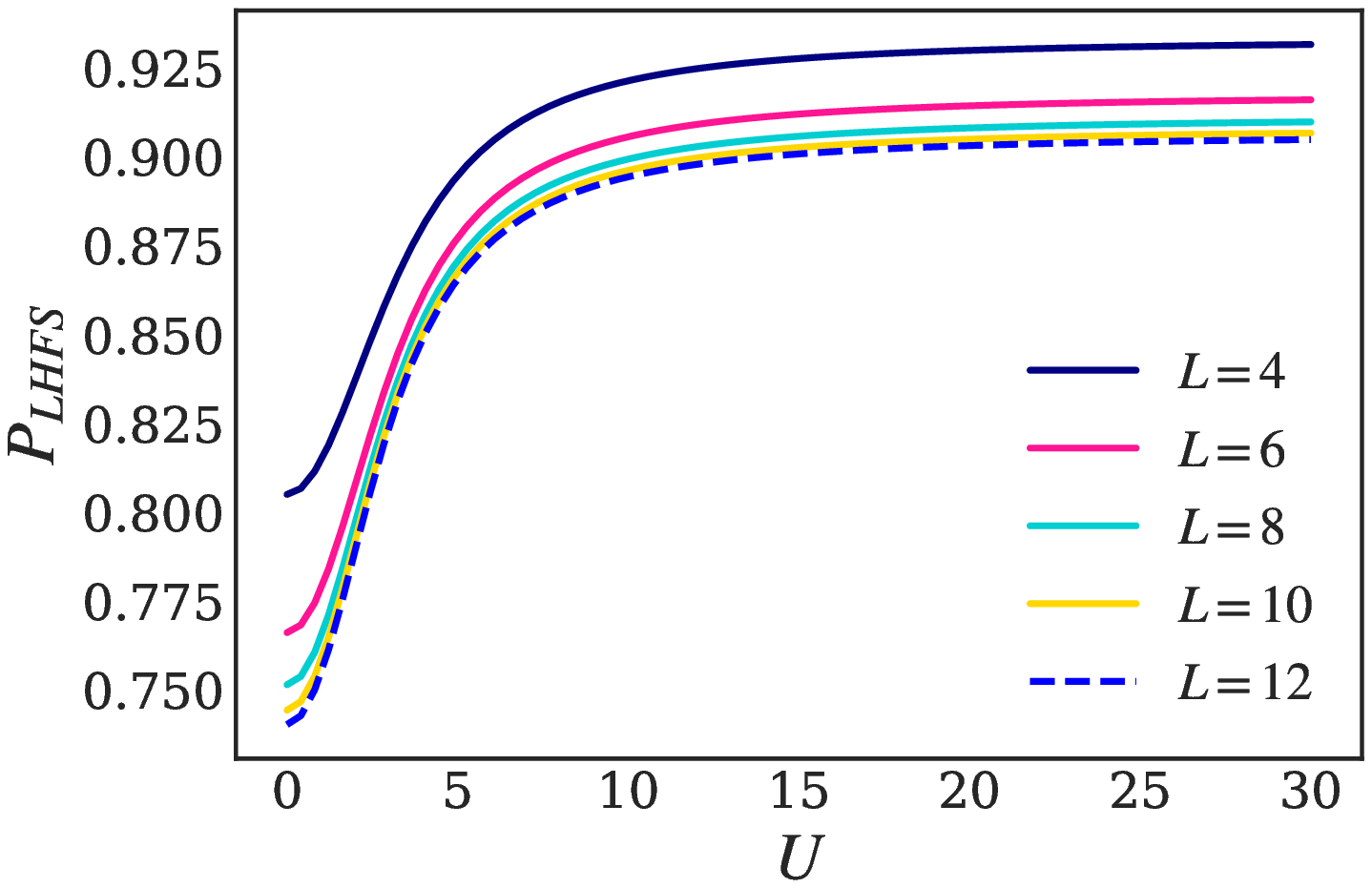}
  \label{fig:img5a}}\quad
  \subfigure[]{\includegraphics[scale=0.35]{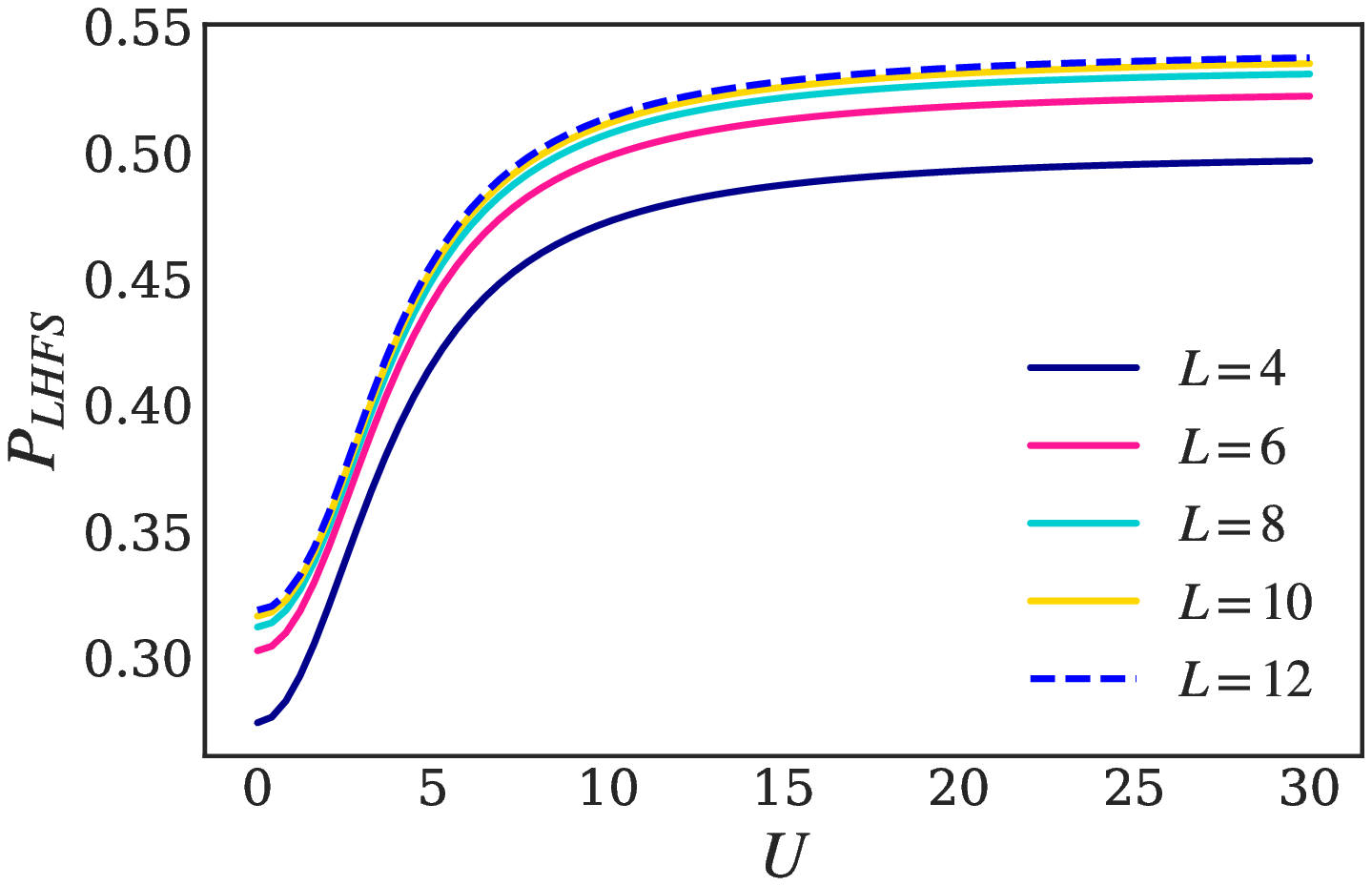}
  \label{fig:img5b}}\quad

  \caption{The probability $P_{k}$ associated to the $LHFS$ as a function of $U$ with various system sizes $L$ for (a) $\rho_{12}$ and (b) $\rho_{23}$ .}
  \label{fig:img5}
\end{figure}

Going back to the size effect on pairwise entanglement, $C(\rho_{ij})$, for an even or odd index $j$, discussed in section \ref{sec: section3}, now the picture is more clarified. Actually, in addition to the aforestated reasoning, the pairwise entanglement, $C(\rho_{ij})$, follows the probability $P_{LHFS}$ for an arbitrary system size, either for even or odd $j$. In FIG.\ref{fig:img5} it is displayed that for an even index $j$, $P_{LHFS}$ decreases with the system size, such is the case for $\rho_{12}$ in FIG.\ref{fig:img5a}, and this is why $C(\rho_{ij})$ decreases too with the increase of the system's size $L$ for $j$ even. However, for an odd index $j$, $P_{LHFS}$ increases instead as $L$ increases, as shown for the pair $\rho_{23}$ in FIG.\ref{fig:img5b} and in this case $C(\rho_{ij})$ increases too with $L$.     

It is appropriate to note that following the reasoning above, the pairwise entanglement of the pair $\rho_{14}$ is greater than $\rho_{13}$, which is indeed the situation observed from FIG.\ref{fig:img1}. At a first glance, this seems unobvious and contradicts the distance effect on pairwise entanglement since it is commonly believed that this later decays rapidly as the distance separating the particles grows, in most systems with short-range interaction. However, an easier explanation for this peculiar behavior can be derived in terms of the entanglement monogamy and CKW inequalities \cite{ckw} through the entanglement that is externally shared by a given pair (we will refer to this type of entanglement as \textit{shared entanglement} hereafter). In fact, since entanglement is actually created as a consequence of the direct interactions and strong correlations appearing in the system's Hamiltonian, a pair that is directly linked with more sites will exhibit a higher of amount of shared entanglement and as a result a smaller amount of pairwise entanglement between the sites composing the pair.

\begin{figure}
  \centering
 
  \subfigure[]{\includegraphics[scale=0.82]{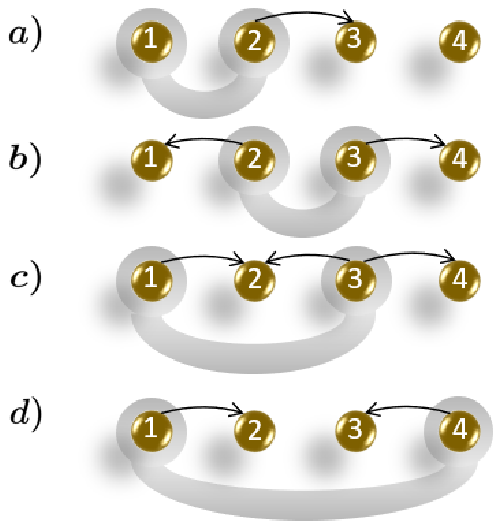}
  \label{fig:img6a}}\quad
  
  \subfigure[]{\includegraphics[scale=0.40]{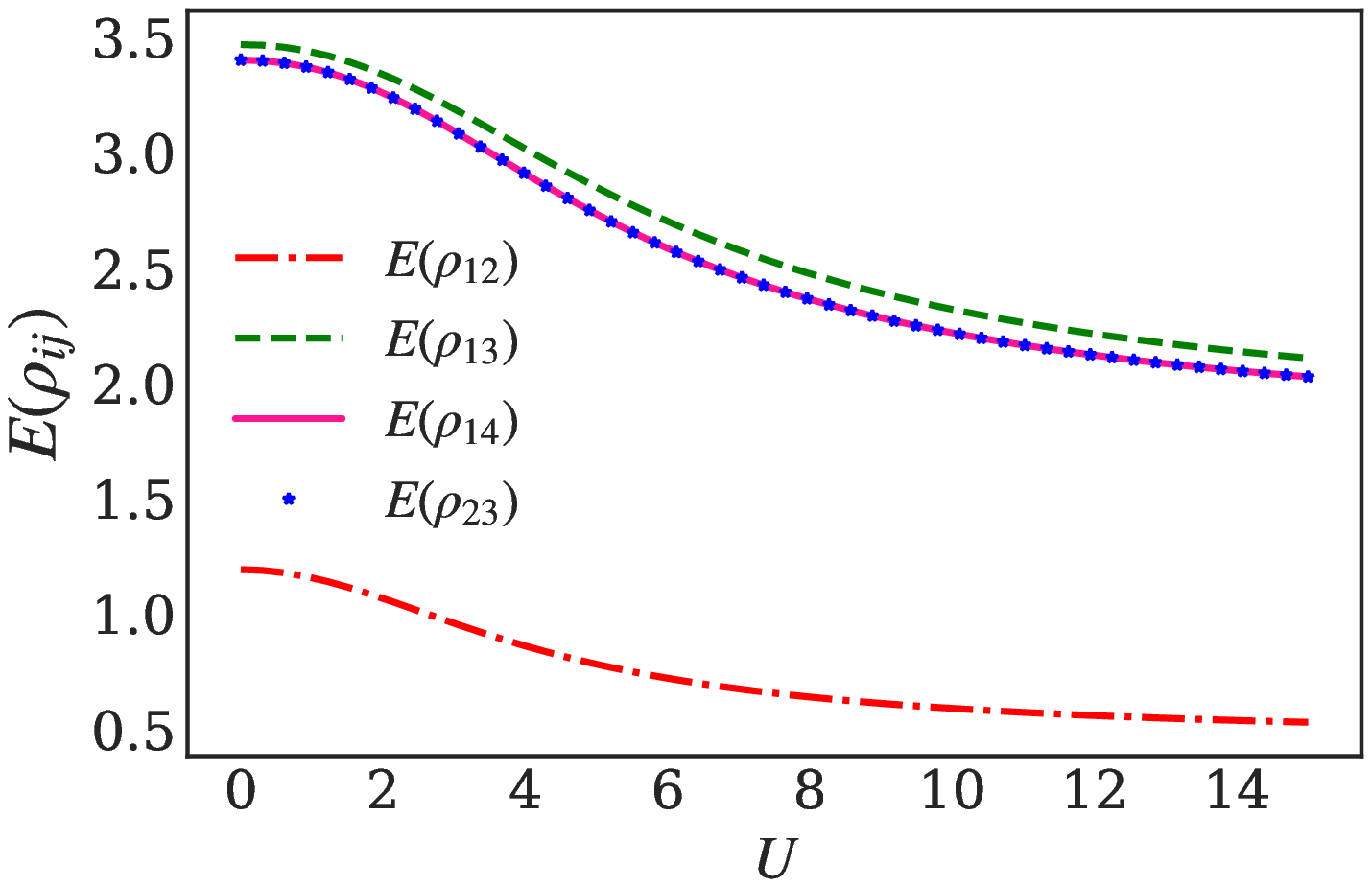}
  \label{fig:img6b}}\quad

  \caption{(a) Representative scheme showing the direct correlation bonds between each pair and the remaining sites for $L=4$. (b) The entanglement $E(\rho_{ij})$ externally shared by the different pairs for $L=4$.}
  \label{fig:img6}
\end{figure}

In this regard, the adjacent sites are strongly correlated compared to the distant sites, because of the rigorous direct interaction present between them, which is generated in the Fermi-Hubbard system by way of the hoping of electrons at $U=0$. So, the pairs $\rho_{12}$ and $\rho_{34}$ (pairs at the ends) show the presence of significant internal correlations (pairwise entanglement) compared to the pair $\rho_{23}$ (middle sites), because as illustrated from FIG.\ref{fig:img6a}, $\rho_{23}$ is directly connected to the sites $1$ and $4$ making two bonds of direct interactions contrary to $\rho_{12}$ which is directly bonded only with the site $3$. In other words, the pair $\rho_{23}$ is able to externally share a great amount of correlations with the other sites in comparison to $\rho_{12}$. This is exactly the point displayed in FIG.\ref{fig:img6b}, where the entanglement $E(\rho_{23})$ shared by the pair $\rho_{23}$ is considerably greater than $E(\rho_{12} )$. As a result, and because of the entanglement monogamy the pairwise entanglement $C(\rho_{12})$ is expected to be substantially higher than $C(\rho_{23})$: $C(\rho_{12} )> C(\rho_{23})$, which is indeed confirmed by FIG.\ref{fig:img1} for $L=4$.

A similar reasoning establishes that $C(\rho_{14})> C(\rho_{13})$. Indeed, in FIG.\ref{fig:img6a} it is plotted that $\rho_{14}$ is directly linked with the sites $2$ and $3$, making two bonds of direct interactions exactly similar to the pair $\rho_{23}$. This is the reason why they externally share the same amount of entanglement as shown in FIG.\ref{fig:img6b}. Furthermore, since they have equal occupation probabilities $P_{k}$ at $U=0$ (FIG.\ref{fig:img4a}), hence $\rho_{23} = \rho_{14}$, which implies that $C(\rho_{23})= C(\rho_{14})$ at $U=0$. This is indeed observed in FIG.\ref{fig:img1b} and FIG.\ref{fig:img1d} for $L=4$. On the contrary the pair $\rho_{13}$ makes tree direct interaction bonds with sites $2$ and $4$ (FIG.\ref{fig:img6a}). For this reason the pair $\rho_{13}$ has to externally share the greatest amount of entanglement $E(\rho_{13})$ compared to all the other pairs as displayed in FIG.\ref{fig:img6b}. Hence, according to the monogamy of entanglement the pair $\rho_{13}$ is less entangled in comparison to $\rho_{14}$ thus establishing the long distance entanglement of the pair $\rho_{14}$ against $\rho_{13}$.

Combining the previous results with the fact that the pairs of adjacent sites are strongly correlated compared to the distant sites and that the adjacent sites at the ends are more entangled than those located at the middle of an open array, allows to set up an order of the different pairwise entanglements observed for $L=4$, as follows 
\begin{equation}
    C(\rho_{12})> C(\rho_{23}) \geq C(\rho_{14}) > C(\rho_{13}).
\end{equation}

The local entanglement, $E(\rho_{i})$ (FIG.(\ref{fig:img3})), as well as the entanglement, $E(\rho_{ij})$, shared by the pairs (Fig.\ref{fig:img6b}), have a maximal value at $U=0$ because of the rich structure of quantum correlations present in the system. Once $U$ increases, the entanglement decreases and here the quantum correlations are no longer powerful. Nonetheless the pairwise entanglement $C(\rho_{ij})$ could achieve a maximal value at $U\ne 0$. This quantum picture will be clarified and studied in details in the next section, where a rigorous explanation of the behavior of pairwise entanglement will be presented.

\subsection{Pairwise entanglement at finite coupling regime}

Generally, increasing $U$ allows the pairs to favor more the LHFS, except for the pair $\rho_{13}$. This latter has a strong tendency to be in a state with $N=2$ but $S=\{-1,0,+1\}$ as a mixture that combines the ferromagnetic and the anti-ferromagnetic behavior associated to this subsystem. It is defined as $\rho=p\sum_{i=1}^{3}\ket{\psi_{i}}\bra{\psi_{i}}$ with $\ket{\psi_{1}}= \ket{\uparrow \uparrow} $, $\ket{\psi_{2}}=\ket{\downarrow \downarrow}$ and $\ket{\psi_{3}}= \alpha \ket{\uparrow \uparrow}+ \beta \ket{\downarrow \downarrow} + \gamma \ket{\uparrow \downarrow} + \delta \ket{\downarrow \uparrow}$. This is clearly shown in FIG.\ref{fig:img4} where the probability associated to the LHFS grows faster with $U$ compared the other states of the pairs $\rho_{12}$, $\rho_{23}$ and $\rho_{14}$, whereas for the pair $\rho_{13}$, it is the probability $p$ associated to $\ket{\psi_{1}}$, $\ket{\psi_{2}}$ and $\ket{\psi_{3}}$, that increases instead and becomes dominant compared to the other probabilities. Evolving into the favored pure LHFS or the mixed state $\rho$, while rising the coulomb interaction $U$, each pair $\rho_{ij}$ exhibits an increase of its purity which is again established from FIG.\ref{fig:img4}, where majority of the probabilities $P_{k}$ vanish. This is among the prominent factors that could increase the pairwise entanglement, as it is widely known that increasing the degree of mixture decreases the entanglement. In our case we observe an increase of the entanglement due to the decrease of the mixture. However, relying solely on this statement is not enough, because as shown in FIG.\ref{fig:img1d} the pairwise entanglement $C(\rho_{14})$ decreases instead; thereby another factor has to be taken into account. As a matter of fact, the state describing the pairs is given by $\rho_{ij}(U)= \sum_{k}P(U) \ket{\psi_{k}(U)}\bra{\psi_{k}(U)}$, in which the states $\ket{\psi_{k}}$, namely the LHFS, evolve also with $U$ at the same time as the probabilities $P_{k}$ evolve. In this regard, we will show that there are two competing effects that are present in the system: on one hand the mixing effect and on the other hand the inherent entanglement effect associated with the LHFS.

\begin{figure*}
  \centering
  \subfigure[]{\includegraphics[scale=0.62]{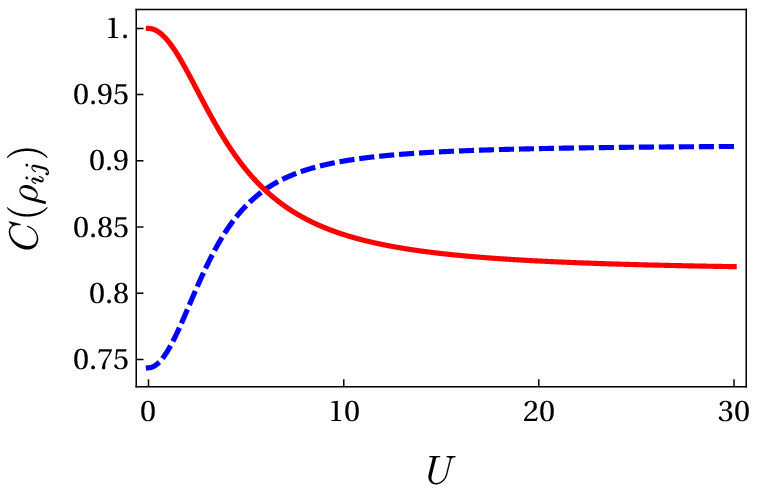}
  \label{fig:img7a}}\quad
  \subfigure[]{\includegraphics[scale=0.60]{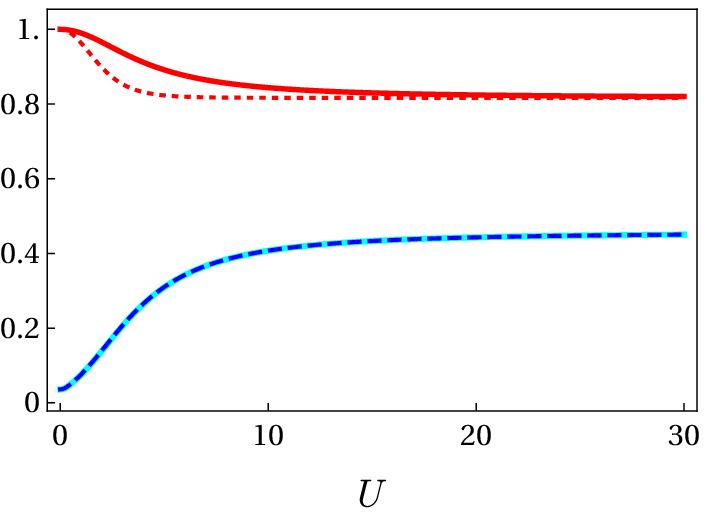}
  \label{fig:img7b}}\quad
  \subfigure[]{\includegraphics[scale=0.60]{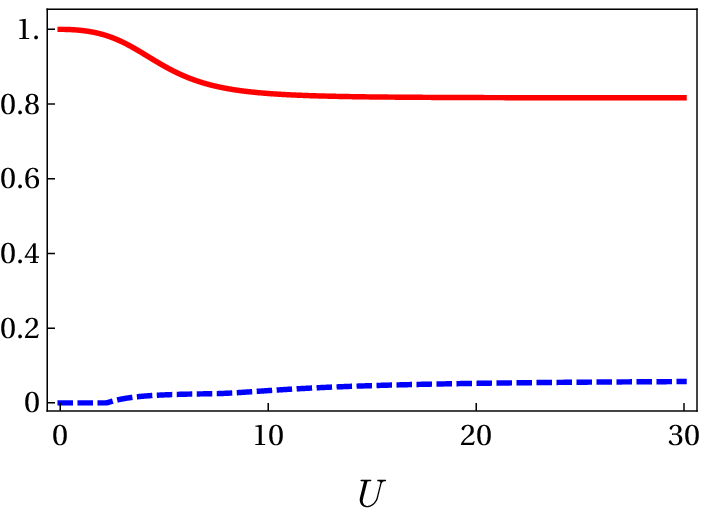}
  \label{fig:img7c}}
  
  \caption{\textbf{Competing effects.} The LBC associated to the LHFS (solid and dotted red lines) and to the global state $\rho_{ij}$ with frozen LHFS (solid and dashed blue lines) for (a) $\rho_{12}$, (b) $\rho_{23}$ and $\rho_{14}$ and (c) $\rho_{13}$.}
  \label{fig:img7}
\end{figure*}

Based on the previous statement let us return to the discussion of the pairwise entanglement behavior with the increase of $U$ as plotted in FIG.\ref{fig:img1}. The aforementioned two effects will play an important role in explaining the behavior of pairwise entanglement. In order to show the effect of the entanglement evolution of the LHFS on the global correlations $C(\rho_{ij})$, we have frozen the evolution of the LHFS with $U$ inside the mixture $\sum_{k} P_{k} \ket{\psi_{k}}\bra{\psi_{k}}$ without freezing the evolution of $P_{LHFS}$ associated to it, such that $\Tr(\rho_{ij}) =1$ is always satisfied. From FIG.\ref{fig:img7} it is clear that for a frozen LHFS, the pairwise entanglement $C(\rho_ {ij})$ increases and then stabilizes for large values of $U$. Apparently, this is due to the decline of the mixing effect shown in FIG.\ref{fig:img4}, where the majority of the probabilities $P_{k}$ vanish with $U$. For a non frozen LHFS, the effect of the entanglement behavior of this latter (displayed in FIG.\ref{fig:img7}) appears after a specific value of $U$ in the plots of the pairwise entanglement $C(\rho_{ij})$, FIG.\ref{fig:img1}, where it is noticeable that $C(\rho_{ij})$ decreases after achieving its maximum. In this case, the inherent effect associated to the LHFS is well dominating the pairwise correlations. 
To be specific, from FIG.\ref{fig:img5}, for small values of $U$, the behavior of $C(\rho_{ij})$ is generally controlled by the decline of the mixing effect resulting in a significant increase of the entanglement in the adjacent sites such as $C(\rho_{12})$ and $C(\rho_{23})$. For the distant sites the increase of entanglement, that appears only in small size systems since the range of entanglement is limited, starts after an interval of $U$ given the fact that the degree of mixedness in this case is higher and thus the mixing effect persists longer. Beyond that, it is the behavior of the LHFS that becomes the dominant one, where the pairwise entanglement $C(\rho_{ij})$ decreases and follows the entanglement behavior of the LHFS itself. At this stage, with the increase of $U$, the system is well focused on the LHFS with a high probability. In contrast, for the pair $\rho_{14}$ with $L=4$ (FIG.\ref{fig:img1d}) the situation is different, where it is clearly shown that the pairwise entanglement $C(\rho_{14})$ decreases instantly with $U$. A further explanation can be more clarified if we look at FIG.\ref{fig:img7b} where the pairs $\rho_{23}$ and $\rho_{14}$ have the same degree of the mixing effect since they have the same probabilities $P_{k}$ as a function of $U$ (FIG.\ref{fig:img4c}). Nonetheless the evolution rate of the LHFS associated to $\rho_{14}$ is considerably faster compared to the LHFS of $\rho_{23}$. This is expected as long as the the correlations at the end sites decay faster as opposed to the middle sites against $U$. Due to this fact, the inherent effect associated to the LHFS becomes the dominant one. In this case the behavior of entanglement will be overwhelmed by the behavior of the LHFS itself and here the decline of the mixing effect is no longer able to rise the pairwise entanglement $C(\rho_{ij})$ \textit{vs} the evolution rate of the LHFS entanglement. Thus the end to end pairwise entanglement generally decreases instantly with $U$, but this behavior is expected to be shown only in small size systems such as $L=2$ and $L=4$, since the long distance entanglement is limited in large size systems.

\section{ \label{sec: section5}Confinement state and genuine 4-partite entanglement}

The Fermi Hubbard sites can be assimilated to quantum dots when $U$ tends to infinity. Actually, when the repulsion interaction $U$ within the sites is strong enough, it allows the electrons to move away from each other (thus excluding the sites with double occupancy) and the tunneling of these electrons between the sites is blocked. The situation is similar to the creation of potential barriers between the sites which prohibits the electrons to move outside. This process displays exactly the quantum confinement effect in the Fermi-Hubbard sites and as such, the state describing the system is called the confinement state with $N=\frac{L}{2}$ and $S=0$.

Apparently, systems with different sizes do behave similarly in most cases. In this respect, the stabilization of pairwise entanglement in the confinement state under the variation of $U$ is a common property characterizing the various system sizes. So following the steps in the previous section, the study of the confinement state will be restricted to the smallest size system $L=4$, then some insights can be drawn for Larger sizes.

For a small size array of four quantum dots, the confinement state (in the strong coupling regime) is given by
\begin{equation}
\label{confstate}
  \begin{aligned}
    \ket{\psi_{C}} = & -\alpha(\ket{ \downarrow  \downarrow  \uparrow\uparrow}+ \ket{\uparrow\uparrow  \downarrow  \downarrow  }) + \beta (\ket{ \downarrow  \uparrow \downarrow \uparrow} +  \ket{ \uparrow \downarrow \uparrow \downarrow  } ) \\ & -\gamma (\ket{ \downarrow \uparrow\uparrow  \downarrow  }+\ket{ \uparrow \downarrow\downarrow  \uparrow}),
  \end{aligned}
\end{equation}
where $\alpha = \frac{1}{\sqrt{6}}\approx 0.41$, $\beta= \frac{3\sqrt{8639}}{500}\approx 0.56$ and $\gamma= \frac{\sqrt{16747/3}}{500}\approx 0.15$.
When $U$ takes strong values the pairwise entanglement associated to the pairs $\rho_{13}$ and $\rho_{14}$ vanishes but for $\rho_{12}$ the entanglement stabilizes with an important value, as seen in FIG.\ref{fig:img1}. It turns out that, when $U\to + \infty$, the pairwise entanglement $C(\rho_{23})$ vanishes too and becomes equal to $C(\rho_{14})$. As the LBC is actually, just an estimation of the minimal value of entanglement, the conclusions that can be drawn from its behavior can be misleading sometimes and as a consequence have to be taken with much care. But in this case, this behavior can be confirmed by returning to the well established measures of entanglement for qubits and using them to get the exact amount of the different pairwise entanglements contained in $ \ket{\psi_{C}}$. As a matter of fact, because each dot can be occupied by one electron with spin $\uparrow$ or $ \downarrow $, the state (\ref{confstate}) can be assimilated now to that of a two-level system. Consequently, the concurrence (for qubits) \cite{ref26} is an appropriate measure of the amounts of pairwise entanglement, and gives in this case the following values $$C( \rho_{12})= 0.866, \; C( \rho_{23})= C( \rho_{13}) = C( \rho_{14}) =0.$$

Going back to the discussion about the shared entanglement, and given the fact that the pair $\rho_{12}$ is still sharing entanglement with the system when $U\to + \infty$ (FIG.\ref{fig:img6b}), this means that other kinds of quantum correlations, beyond pairwise entanglement, are present. Namely, the genuine four-partite entanglement. It turns out that the confinement state $\ket{\psi_{C}}$ belongs to the generic class that represents one classification of the nine families of four-qubits' pure states defined in \cite{ref27}. It can be generally written in the computational basis as follows
\begin{equation}
 \small
  \begin{aligned}
   \ket{\psi_{G}} = \frac{z_{0} + z_{3}}{2}(\ket{0000}+\ket{1111})+ \frac{z_{0} - z_{3}}{2} (\ket{0011}+\ket{1100})  \\ + \frac{z_{1} + z_{2}}{2}(\ket{0101}+ \ket{1010}) + \frac{z_{1} - z_{2}}{2} (\ket{0110}+ \ket{ 1001}),
   \end{aligned}
 \normalsize
\end{equation}
with $z_{0},z_{1}, z_{2}, z_{3} \in \mathbb{C}$. For the confinement state $\ket{\psi_{C}}$ in (\ref{confstate}), $z_{0}=-z_{3} = -\alpha$, $z_{1}= \beta-\gamma$ and $z_{2}= \beta+\gamma$. The four tangle for such state can be defined and computed as \cite{ref28,ref29,ref30} $\tau_{1234}^{(\psi_{C})} = \lvert \sum_{i=0}^{3} z_{i}^{2} \rvert^2 =\lvert 2(\alpha^{2} + \beta^{2} + \gamma^{2}) \rvert^2 = 1$ with $i=0,1,2,3$. So the confinement state is maximally entangled when it comes to the genuine four-partite entanglement.

It is worth noting that the four-qubits $GHZ$ state belongs also to the same generic family with $z_{0}=z_{3}=1/\sqrt{2}$ and $z_{1}=z_{2}=0$ where we have $\tau_{1234}^{GHZ} = \lvert 2z_{0}^{2}\rvert^2 = 1$. 

Thus the Hubbard model's confinement state is genuinely multi-partite entangled even for larger sizes. An n-partite entangled state is called genuine if and only if the state is not separable with respect to any m-partition ($m\leq n$) and this is confirmed by the fact that entanglement $E(\rho_{m/n})$ in the Hubbard model has a non-zero value. However, defining the family class in which the confinement state, as a n-qubit state, is belonging is still an interesting question to be answered in future works and seems to get complicated as the system size becomes larger $L>4$.

\section{\label{sec: section7} Conclusion and perspectives}

In this work we have examined the pairwise entanglement in quantum dot systems, formally described by the one dimensional Fermi-Hubbard model. As ququart systems, the lower bound of concurrence is deemed the appropriate choice to prospect on the behavior in the weak coupling regime. Namely, it is demonstrated that a proper optimization of the Coulomb interaction can create entanglement between quantum dots and even more so, help it to grow considerably. On a related note, it was shown that the range of entanglement is well extended to the third neighboring site for a system size of $L=4$, whereas it could be created and extended to the third neighboring site by properly adjusting the Coulomb interaction  for $L>4$. 

In addition to that, the size effect was studied and it was shown that under the size effect the pairwise entanglement decreases for the pairs $\rho_{ij}$ with $j$ even as the size grows, whereas it increases if $j$ is odd. In this regard, we have presented a rigorous description of the pairs in terms of the local half filled state for each pair with a fixed number of electrons $N=2$ and a spin $S=0$. Acquaintance with this state provides a proper explanation concerning the amount of pairwise entanglement as well as the behavior of this latter against the Coulomb interaction and the size effect. On a last note, the study of the confinement state in the system demonstrated the existence of the genuine four-partite entanglement with a maximum value achieved for $L=4$. 

Motivated by these results, it would be of interest to study the genuine multi-partite entanglement as well as the family class in which the confinement state is belonging for larger sizes. Furthermore it is clear that the Hubbard model provides an adequate entanglement resource for ququart teleportation, however as the size increases the end to end entanglement vanishes which allows us to study in future works some mechanisms to circumvent this by producing long range entanglement in the Fermi Hubbard model allowing thus to reach maximum teleportation fidelity.

\begin{acknowledgments}
S. A. acknowledges gratefully the National Center for Scientific and Technical Research (CNRST) for financial support (Grant No. 1UM5R2018). This research was supported through computational resources of HPC-MARWAN (www.marwan.ma/hpc) provided by CNRST, Rabat, Morocco. The calculations were done using QuSpin \cite{ref31,ref32}. The authors acknowledge fruitful discussions with Z. Mzaouali.

\end{acknowledgments}



\nocite{*}

\bibliography{apssamp}

\end{document}